\documentclass[10pt,conference]{IEEEtran}
\pdfoutput=1
\usepackage{cite}

\usepackage{amsmath,amssymb,amsfonts}
\usepackage{algorithmic}
\usepackage{graphicx}
\usepackage{textcomp}
\usepackage[dvipsnames]{xcolor}
\usepackage[colorlinks=false,
citebordercolor={.85 1 .85},
linkbordercolor	={1 .85 .85},
urlbordercolor	={.85 1 1}
]{hyperref}
\usepackage{amssymb}
\usepackage{MnSymbol}
\usepackage{wrapfig}
\usepackage{array}
\usepackage{bm}
\usepackage{graphicx}
\usepackage{float}
\usepackage{url}
\usepackage{enumitem}
\usepackage{adjustbox}

\usepackage{accents}
\usepackage{lipsum}

\DeclareMathOperator*{\argmax}{arg\,max} 
\usepackage[all,cmtip]{xy}

\usepackage[braket]{qcircuit}

\newcommand{\qcscaled}[2]{\scalebox{#1}{\Qcircuit @C=0.3em @R=1em {#2}}}

\newcommand{\bigO}{\mathcal{O}}

\newcommand{\td}[1]{}     
\newcommand{\tdok}[1]{}  
\newcommand{\jk}[1]{}  

\makeatletter
\def\widebreve{\mathpalette\wide@breve}
\def\wide@breve#1#2{\sbox\z@{$#1#2$}%
	\mathop{\vbox{\m@th\ialign{##\crcr
				\kern0.08em\brevefill#1{0.8\wd\z@}\crcr\noalign{\nointerlineskip}%
				$\hss#1#2\hss$\crcr}}}\limits}
\def\brevefill#1#2{$\m@th\sbox\tw@{$#1($}%
	\hss\resizebox{#2}{\wd\tw@}{\rotatebox[origin=c]{90}{\upshape(}}\hss$}
\makeatletter

\begin{document}
\bstctlcite{IEEEexample:BSTcontrol}
	\title{Polyadic Quantum Classifier}

	\author{\IEEEauthorblockN{William Cappelletti, Rebecca Erbanni and Joaqu\'{i}n Keller }
		\IEEEauthorblockA{\textit{Entropica Labs},
			Singapore \\
			\{william, rebecca, joaquin\}@entropicalabs.com }
	}

	\maketitle

\begin{abstract}
	We introduce here a supervised quantum machine learning algorithm for multi-class classification on NISQ architectures. A parametric quantum circuit is trained to output a specific bit string corresponding to the class of the input datapoint.

	We train and test it on an IBMq 5-qubit quantum computer and the algorithm shows good accuracy ---compared to a classical machine learning model--- for ternary classification of the Iris dataset and an extension of the XOR problem.

	Furthermore, we evaluate with simulations how the algorithm fares for a binary and a quaternary classification on resp. a known binary dataset and a synthetic dataset.
\end{abstract}


\nocite{daskin2018iris,grant2018,schuld2017iris,farhiNeven2018,schuld2020circuitcentric,lloyd2020qml,latorre2020,romero2019variational,multiclass2005survey}

\section{Introduction}
Quantum machine learning (QML) has raised great expectations, it is thought\cite{nisq2018}\cite{tfquantum2020} to be one of the  first possible applications of quantum computing to be able to run on NISQ\footnote{Noisy Intermediate-Scale Quantum} computers.

Nonetheless, QML is still in its infancy. We can make a parallel with the dawn of machine learning in the 1950s when the emblematic perceptron \cite{perceptron1958, mazieres2018revanche} was introduced to solve binary classification problems. Today's research in QML has followed the same path and  binary classification has been broadly studied.  

In 1969\cite{minsky1969} it was shown that the perceptron could not solve the simple XOR problem. In fact it can only classify linearly separable datasets and it wasn't before 1986 that multilayer perceptrons with backpropagation \cite{backprop1986} addressed harder problems.
For instance, the ternary classification of the Iris flower dataset\cite{fisher1936iris}, which is non-linearly separable, could not be solved with the perceptron approach, but is now a typical test case\cite[Ch. 2]{ESL} of machine learning.

In this paper, we introduce a QML algorithm for multi-class classification and challenge it with the Iris flower dataset and a extension of the XOR problem with added Gaussian noise. The Iris flower dataset has three classes, two of which are not linearly separable. Binary QML classifications on this dataset have been addressed, on the linearly separable class against the rest, by \cite{daskin2018iris, grant2018} and, pairwise, on all classes by \cite{schuld2017iris}.

Any n-ary classifier can be implemented with n binary classifiers\cite{multiclass2005survey} however, to predict all classes at once, we take a direct approach described in section \ref{section:output}.

There was no guarantee it would be possible to train our classifier on an actual quantum computer. However our simulations of the algorithm on the Iris dataset show that the quantum computing power needed, for both training and test, can be found in today's hardware. And since models trained on simulators were tested on IBMq by \cite{schuld2017iris} and \cite{grant2018}, we were optimist that the inherent noise of the hardware wouldn't be an impediment. Indeed, we run our algorithm ---train and test--- on IBMq quantum hardware, for the ternary classification of the Iris dataset. Results are in section~\ref{ssec:iris-exp}.

Similarly, in section~\ref{ssec:xor-exp}, we successfully trained a model for the Gaussian XOR problem on the same quantum system.
%

Other experiments, with a simulator, on
binary and quaternary classifications suggest that our approach is flexible enough to be applied to many problems. The outcome of these experiments are in sections \ref{ssec:skin-exp} and \ref{ssec:synthetic}.

As detailed in sections \ref{section:vqa} and \ref{section:input}, our algorithm is based on parametric quantum circuits. Our approach is mostly empirical and the exact design of these circuits relies on experiments and on some guidelines presented in sections \ref{section:optimizing} and \ref{section:rules}.

		\section{A variational quantum algorithm}
	\label{section:vqa}

	Like many quantum machine learning algorithms\cite{daskin2018iris,grant2018,schuld2017iris,farhiNeven2018,schuld2020circuitcentric,lloyd2020qml,latorre2020,romero2019variational},
	our procedure for classification of classical data fits in the general scheme of variational quantum algorithms\cite{peruzzo2014vqa},
	where a parametric quantum computation $\mathfrak{F}_{\bm{\theta}}$ is applied to an input vector $\bm{x}$ to get the result $\hat{y}=\mathfrak{F}_{\bm{\theta}}(\bm{x})$.

	The variational algorithm per se consists of running $\mathfrak{F}_{\bm{\theta}}(\bm{x})$ for different values of $\bm{\theta}$ and $\bm{x}$ to find an optimal $\bm{\theta}^\star$
	for which the results are satisfactory.

	Although some have envisioned hardware architectures with quantum random access memory \cite{qram2008}
	and other interesting features, our algorithm relies only on the simplest functionalities available in actual quantum hardware. \jk{This point on actual hardware needs more emphasis}

	\newcommand{\qp}{{\mathtt{P}}}

	In most of today's quantum computers or QPUs\footnote{Quantum Processing Unit}, an elemental computation $\mathtt{Q}$ take as input $n$,  a number of \emph{shots}, and $\qp$, a quantum program, or circuit.  A circuit $\qp$ is  a sequence of \emph{gates}, or elemental operations on qubits. 

	In what is called a shot, the qubits of the quantum computer are all initialized at $\ket{0}$, the sequence $\qp$ of gates is applied to the qubits, then they are all measured. A \emph{run} or \emph{circuit run} is a sequence of $n$ shots.  Runs and shots are quantum computations of different granularity. The result  $\mathtt{Q}\left(n,\qp\right)$ of a circuit run is the sequence $\hat{R}$ of $n$ bit strings in $\{0,1\}^N$, 
	 corresponding to the measurement of the $N$ qubits of the circuit.

	 We can note that, the computational time complexity of $\mathtt{Q}\left(n,\qp\right)$ is $\bigO(n\times |\qp|)$ where $|\qp|$ is the number of gates in $\qp$.

	\newcommand{\mymultigateC}{\qp_{\!\bm{\omega},\bm{\theta}}}
\newcommand{\bgamma}{{\bm{\omega}}}

\section{How to Input Data}
\label{section:input}
To input data we resort to a method called variational, or parametric, encoding which was first proposed in \cite{romero2019variational}.
The method consists of using a parametric circuit to encode the input vector $\bm{x}$ to parameters of the circuit.

A parametric circuit is a circuit where some gates can take continuous angles as parameters and are $2\pi$-periodic regarding  them.

We define an encoding function $f$ to map each coordinate or \emph{feature}, of the input vector $\bm{x}$ to an angle in the interval $\left]-\pi,\pi\right[$, which gives us a vector of parameters $\bm{\omega}=f(\bm{x})$ to be used in a parametric circuit $\qp_{\!\bm{\omega}}$.

\begin{wrapfigure}{r}{0.45\columnwidth}
	\centering
	\qcscaled{0.9}{
		&\lstick{\ket{0}}&\qw&\qw&\qw& \multigate{2}{\mymultigateC}&\qw&\qw&\meter\\
		&\cdots&&&& \nghost{\mymultigateC}&&&\cdots&&& \\
		&\lstick{\ket{0}}&\qw&\qw& \qw&\ghost{\mymultigateC}&\qw&\qw&\meter
	}

	\caption{The input vector $\bm{x}$ is encoded as angles $\bm{\omega}=\!\!f\!\left(\bm{x}\right)$ of a parametric circuit $\qp_{\bm{\!\omega},\bm{\theta}}$ }
\end{wrapfigure}

We note $\bm{X}$ the set of input vectors $\bm{x}$ from the learning dataset and define the vectors $\bm{\overline{X}}$ and  $\sigma_{\bm{X}}$, as its element-wise mean and standard deviation. Similarly, we compute the element-wise standard score $\bm{z}_{\bm{X}}(\bm{x})=\frac{\bm{x}-\bm{\overline{X}}}{\sigma_{\bm{X}}}$,

In a Gaussian distribution approximation, we define the quantile $q=\Phi^{-1}(1-\epsilon^{\frac{1}{d}}/2)$, where $d$ is the dimension of $\bm{X}$ and $\Phi^{-1}$ the quantile function\cite{wasserman2013all}.
By definition, the points $\bm{x}$ such that $\exists i\; \left|\bm{z}_{\bm{X}}(\bm{x})\right|_i>q$  represent less than an $\epsilon$ fraction of $\bm{X}$.
\jk{Emmanuel Viennet: it is a preprocessing routinely used in ML and should not have to be detailed here ? (== « standardization of the data »} %
 We fix $\epsilon$ small and ignore such points.

Thus, we define the encoding function as a simple linear rescale and shift of the input:
$$ f(\bm{x}) = \left(1 - \frac{\alpha}{2}\right) \frac{\pi}{q} \:\bm{z}_{\bm{X}(\bm{x})} . $$
By definition, all angles of the encoded vector $\bm{\omega} = f(\bm{x})$ fall within the interval $\left]-\left(1 - \frac{\alpha}{2}\right) \pi,\left(1 - \frac{\alpha}{2}\right) \pi\right[$.
This enforces an angular gap $\alpha \pi$ between the extreme values of the encoded dataset, where $\alpha$ is a parameter to be choosen.

Using this encoding function $f$, we can now
define our quantum classifier as
$$\hat{y}=\mathfrak{F}_{\bm{\theta}}(\bm{x})= g\left(\;\mathtt{Q}\left(n,\qp_{\!\bm{\omega},\bm{\theta}}\right)\;\right) , $$
where $\qp_{\!\bm{\omega},\bm{\theta}}$ is a parametric quantum circuit,  $\bm{\omega}=f(\bm{x})$ are the encoded input parameters and $\bm{\theta}$ the model parameters,
i.e.\ those to optimize.  The parameter $n$ is the number of shots, which here is not automatically learned but adjusted ``by hand"; and $g$ is the postprocessing function, a classical computation ---described in following sections.

We can note that as $f$, the input encoding function,  is not parametric, the dataset is encoded only once, prior to the learning process.

\newcommand{\commentblock}[1]{}
\commentblock{
However, to input data without loss of information, the inputing process into a quantum state needs  to be injective.
Here the quantum end
$h_\psi(\bgamma) = U_\bgamma \ket{\psi}$ where $U_\bgamma$ is the unitary of $\qp_\bgamma$.
If we make abstraction of noise in the operation of the quantum hardware, this condition is satisfied iff there is a function\footnote{The function $U^{-1}$ only needs to have a theoretical existence and do not need to be effectively calculable.}
    $U^{-1}$ that from an encoding quantum state $\psi_\bgamma=U_\bgamma \ket{0^N}$ can get back to the original encoded value $\bgamma=U^{-1}(\psi_\bgamma)$. \td{There is an issue with the equality. It seems you are assuming that $U^{-1}$ acts as a map on $\psi_\bgamma$, but $\psi_\bgamma$ is a quantum state so $\gamma$ should be a quantum state too. This needs to be defined/abstracted better than it is shown now}
    When a parametric circuit $U_\bgamma$ fulfill this condition, i.e.\ when from $\ket{0^N}$ it can encode all the parameters $\bgamma$ without loss of information, we will say that this parametric circuit is \textit{invertible} regarding $\bgamma$ or just \textit{invertible} when no confusion is possible.
}
    \tdok{You are trying to say something important, but this definition is sloppy. If the encoding is unitary then the operation is always invertible. However here you are using $U$ to define a circuit (a very bad idea), which makes for great confusion. If you want to offer a more general description, then you need to describe the encoding procedure as a quantum channel or a CPTP map. In alternative, you need to expand the description of the quantum circuit in the previous section and clarify that the operation is unitary in theory, but it might not be in practice - and you have to be very careful when dealing with that.}
    \jk{very good comment, this thing about `invertible' is not useful in this paper, so I removed it. However, what needs to be invertible is not the unitary but this function $h_\psi(\bgamma) = U_\bgamma \ket{\psi}$, invertible on $\bgamma$, the initial state $\psi$ is the parameter and $\bgamma$ and the variable: meaning if you know $\psi$ and the resulting state $U_\bgamma \ket{\psi}$ you can get $\bgamma$. Badly explained + useless $\implies$ removed }

\section {How to Output Data}
\label{section:output}
The output of a basic quantum computation, the run of a circuit $\qp_{\bm{\omega},\bm{\theta}}$, is a sequence $\mathtt{Q}\left(n,\qp_{\bm{\omega},\bm{\theta}}\right)=\hat{R}$ of $n$ bit strings in $\{0,1\}^N$ where $N$ is the number of quantum bits and $n$ the number of shots.

The outcome of each measurement is a bit string $s$ and from quantum mechanics we know that  it follows an underlying probability distribution $P(s)$.
We can estimate its probability by ${\hat{P}(s)=\hat{C}(s)/n}$, where $\hat{C}(s)$ is the number of occurrences of $s$ in $\hat{R}$.

Here, recalling that $\bm{\omega}$ encodes for the input vector $\bm{x}$, we use this estimated probability $\hat{P}(s)$ to predict $\hat{y}$ the class of $\bm{x}$.
In order to do so, we associate to each possible class $k$ a bit string $s_k$ in $\{0,1\}^N$ and we note $\hat{P}(s_k)$ as $\hat{P}_k$.

The output of our algorithm, the predicted class is
$$\hat{y}=\mathfrak{F}_{\bm{\theta}}(\bm{x}) =g\left(\:\mathtt{Q}\left(n,\qp_{\bm{\omega},\bm{\theta}}\right)\: \right) =\argmax_{k}\hat{P}_k . $$
Or, saying it otherwise, the predicted class $\hat{y}$ is the class $k$ for which its associated bit string $s_k$ has the highest number of occurrences. 

Note that, even though $\hat{P}_k$ varies from one run to another, if the relative difference  between the theoretical $P_k$ associated with each class is high enough, $\argmax_{k}$ will return a consistent value even with a low number of shots.

This approach is by essence polyadic, the number of classes being bounded by the total number of possible bit strings. As \cite{latorre2020}, it does not rely on multiple binary classifications\cite{multiclass2005survey}.

\section{Learning}
\label{section:learning}
The learning process for our classifier consists in finding a set of parameters $\bm{\theta}^\star$ for which the predictor ${g\left(\:\mathtt{Q}\left(n,U_{f(\bm{x}),\bm{\theta}^\star}\right)\: \right)}$ has high accuracy. 

To do so, for each sample $(\bm{x},y)$ of the learning dataset $\mathcal{T}$, we define an adequate loss function which correlates with the error of the predictor:
$$\mathcal{L}_{\bm{\theta}}(\bm{x},y)=-\log {\frac{e^{\hat{P}_y}}{\sum_{k\in\mathit{K}}e^{\hat{P}_k}}} , $$
where $\hat{P}_y$ is the probability estimate of the bitstring associated with the real class $y$ of $\bm{x}$.

The optimization process consists of iteratively exploring the space of parameters $\bm{\theta}$ with an heuristic. At each step $i$ we compute the loss function $\mathcal{L}_{\bm{\theta}^i}$ as the average over a random subset $\mathcal{B}_i$ of  $\mathcal{T}$:
$$\mathcal{L}_{\bm{\theta}^i} = \sum_{(\bm{x},y)\in\mathcal{B}_i\subset\mathcal{T}}\frac{\mathcal{L}_{\bm{\theta}^i}(\bm{x},y)}{\scriptstyle{|\mathcal{B}_i|}} . $$
The subset, or \emph{minibatch}, $\mathcal{B}_i$ could be a singleton, the full training set $\mathcal{T}$ or have any cardinality in between.

The result of the training, the vector of parameters $\bm{\theta}^\star$, is the ${\bm{\theta}^i}$ that gives the lowest value for the loss function.
	\newcommand{\uqp}{{\mathtt{\breve{P}}}}
\newcommand{\ketzero}{\ket{0}&&}

\newcommand{\cz}{C\!z}
\section{Optimizing Parametric Quantum Circuits}
\label{section:optimizing}
To specify our circuits we choose an universal set of quantum gates: $s_x^i$  ---the $\pi/2$ rotation about the Pauli-X axis, $Z^i_{\phi}$ ---a rotation of an arbitrary angle $\phi$  about the Pauli-Z axis, and $\cz^i_j$ ---the controlled-$Z$ gate, where $i$ and $j$ are the qubits to which they apply.

In a QPU, these elemental gates are implemented using pulses ---physical phenomena that modify the state of the qubits. A controlled-$Z$ gate is translated in a 2-qubit pulse, a $s_x$ gate needs a 1-qubit pulse and $Z$-rotations need no pulse\cite{zgates2016virtual} as they are implemented by modifying the subsequent pulses.

Each QPU might have a slightly different set of elemental gates, by choice of the hardware manufacturer. Nonetheless, a circuit written using our set of elemental gates have a translation that preserves the number of 1-qubit pulses and 2-qubit pulses for most of the existing QPUs (see Appendix \ref{native}).

A quantum program, or, circuit $\qp$ specifies\cite{nielsen00} a unitary matrix, that we note here $\uqp$. The unitary $\uqp$ transforms a quantum state into another one. 

\newcommand{\mm}{\mathcal{M}}

Qubits in a given quantum state $\psi$ yield, when measured, a bit string $s$ with probability $P(s) = \left|\:\psi_s\:\right|^2$, where $\psi_s$ is the amplitude associated with $s$.
We note $\mm$ the measurement operator $\mm : \psi \mapsto s$.
From the definition of $\mm$ it follows that
  $$ \exists \alpha\!\in\!\mathbb{R}\ \ U\!=e^{i\alpha}U' \quad\implies\quad  \mm U = \mm U'. $$
If this holds we say that $U$ and $U'$ are \emph{equal up to a global phase} and note $U\equiv U'$.

The purpose of a quantum computer is to implement the unitary $\uqp$ and the final measure operation.
However, due to engineering limitations, the hardware is imperfect and the actual quantum operations differs from those specified in $\qp$.
This difference or \emph{noise} grows with the number of pulses and the time needed to run the circuit.

\begin{wrapfigure}{r}{0.45\columnwidth}
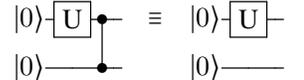

	\centering
		\qcscaled{1.05}{\ketzero&\gate{\mbox{U}}&\ctrl{0}\qwx[1]&\qw&\qw&&&&\equiv&&&&&&\ketzero&\gate{\mbox{U}}&\qw&\qw \\
			\ketzero&\qw&\ctrl{0}&\qw&\qw&&&&&&&&&&\ketzero&\qw&\qw&\qw
		}

	\caption{\label{fig:nocz} A controlled-$Z$ after initialization has no effect, see \ref{nocz}.}
\end{wrapfigure}

Hence, to minimize the noise, if two circuits $\qp$ and $\qp'$ specify for equivalent unitaries $\uqp \equiv \uqp'$ or yield the same result $\mm\uqp = \mm\uqp'$  we will choose the circuit minimizing the number of pulses.

We can now list some simple properties on the unitaries of our elemental gates that will help in optimizing our circuits. Since the unitary ${\widebreve{AB}}$ of the concatenation $AB$ of quantum programs $A$ and $B$ is the composition of their unitaries $\breve{A}\breve{B}$ we will omit the $\breve{\, .\, }$ symbol when no confusion is possible.

\newcommand{\ketkn}{{\ket{0}^{\otimes N}}}

\begin{enumerate}[label={(\arabic*)}]

	\item \label{zz}As $Z^i_\phi Z^i_\lambda \equiv Z^i_{\phi+\lambda}$, in a circuit, two consecutive Z-rotations on the same qubit can be replaced by just one.

	\item \label{zeroz}
		As $Z^i_\phi \ketkn \equiv \ketkn$, a Z-rotation immediately after qubit initialization can removed.
	\item \label{nocz} For any 1-qubit unitary $U^i$ applied to qubit $i$, as $\cz^i_j U^i\ketkn \equiv U^i\ketkn$, we can remove a $\cz$ applied to a qubit $j$ immediately after its initialization. See Fig.~\ref{fig:nocz}.
	\item \label{mz}As $\mm Z_\phi U = \mm U$, we can remove a Z-rotation applied to a qubit just before its measurement.
	\item Similarly, as $\mm \cz^i_j U = \mm U$, we can remove any controlled-Z  followed by no gate before measurement.
	\item \label{canon}As $\cz^i_j\cz^i_j$ is equal to the identity, two consecutive controlled-Z on the same qubits can be removed.
	\item \label{unitary}Any unitary $U^i$ on one qubit $i$ can be decomposed as $U^i\equiv Z^i_{\phi} s^i_x Z^i_{\alpha} s^i_x Z^i_{\lambda}$.
	\item \label{begin} Properties  \ref{unitary}, \ref{zeroz}, \ref{canon} and \ref{zz} imply that any qubit should have at most $s_xZ_\phi s_x$ between initialization and its first $\cz$ gate.
	\item \label{middle} Properties \ref{unitary}, \ref{canon} and \ref{zz} imply that any qubit should have at most a sequence $s_xZ_\phi s_xZ_\alpha$ between two $\cz$ gates
	\item \label{end}Properties \ref{unitary}, \ref{mz} and \ref{zz} imply that any qubit should have at most $s_xZ_\phi s_xZ_\alpha$ after its last $\cz$ gate and before measurement.
\end{enumerate}

\newcommand{\SX}{\gate{\color{MidnightBlue}s_x}}
\newcommand{\zphi}[1]{\gate{Z_{\phi_{#1}}}}
\newcommand{\zgate}[2]{\gate{\color{Mahogany}Z_{{#1}_{#2}}}}
\newcommand{\zzgate}[1]{\zgate{\alpha}{#1}&\SX&\zgate{\phi}{#1}&\SX}

\newcommand{\zphiphi}[2]{\zphi{#1}&\SX&\zphi{#2}&\SX}

\begin{figure}[H]
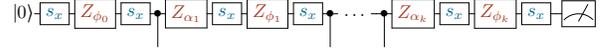

	\centering
	\qcscaled{.7}{
		\ket0&&&\SX&\zgate{\phi}{0}&\SX&\ctrl{1}&
		\zzgate{1}&\ctrl{1}&
		\qw&&& {\dots}&&&& \qw& \ctrl{1} &
		\zzgate{k}& \qw& \meter \\
		&&&&&& &&&&&& &&&&&& &&&&&&
		}
	\caption{\label{maxatomix}Maximum succession of atomic gates on a single quantum bit.}
\end{figure}

Properties \ref{begin}, \ref{middle} and \ref{end} give a maximal sequence
Any circuit with more gates than that can be optimized further.

	\section {Designing circuits}
\label{section:rules}

The constraints for optimality leave a high level of freedom on how to design parametric circuits: less elemental gates are still possible; some angles can be fixed as constants; some parameters will encode input features, while others will be parameters to optimize, and a choice has to be made on how to entangle qubits.

Following intuitions and empirical evidences (some circuits perform better than others) we have identified a set of rules to design what we think are good parametric circuits for our algorithm.
However as intuitions are often misguided and empirical evidences can be overthrown by new experiments, we hope and expect these rules to be challenged and/or expanded by further research work.

\begin{wrapfigure}{r}{0.5\columnwidth}
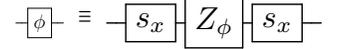

	\centering
		\qcscaled{.75}{&\qw&\gate{\phi}&\qw&\qw}  $\:\equiv\:$  \qcscaled{1.2}{&\qw&\gate{s_x}&\gate{Z_\phi}&\gate{s_x}&\qw&\qw}

	\caption{\label{fig:notation} Compact notation for $s_xZ_\phi s_x$.}
\end{wrapfigure}

Two main criteria enter in the design of parametric circuits intended to be used as classifiers.
First, we want to minimize the number of gates.
Second, we need enough parametric gates to encode the input vector and provide adequate learning capacity.

\begin{figure}[H]
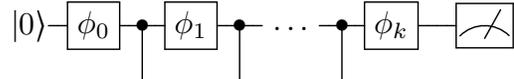

	\centering
	\qcscaled{1.2}{
		\ket0&&&\qw&\gate{\phi_0}&\qw&\ctrl{1}&\qw& \gate{\phi_1}&\qw&\ctrl{1}&\qw&\qw&&& {\dots}&&&& \qw&\ctrl{1}&\qw& \gate{\phi_k}&\qw&\qw& \qw& \meter \\
		&&&&&& &&&&&& &&&&&& &&&&&&
	}
	\caption{\label{fig:succession}Succession of operations on a quantum bit, only one parametric gate between entanglements. }
\end{figure}

This double constraint --minimize the number of gates and maximize the number of parametric gates-- should have led to keep all the $Z$-rotations as parametric gates.
Yet, without fully understanding why, we noticed that a parametric $Z$-rotation right after a controlled-$Z$ seem not to add much capacity and tend to impair the learning phase. Hence, we opt for a rule enforcing exactly one parametric gate between entanglements, or more precisely, as shown in  Fig. \ref{fig:succession}, a sequence $s_xZ_\phi s_x$ between any two consecutive controlled-$Z$ gates.

If we extend this rule for individual qubits to the whole set of qubits, we have a step where each qubit is rotated by some angle followed by a step where qubits are entangled by pairs. Things are slightly, but not fundamentally, different depending whether the number of qubits is odd or even.

Although we don't have clear rules on how or why choose a given entangling pattern at each step, it is well understood~\cite{jozsa2003} that to unleash the power on quantum computing qubits need to be highly entangled.

At each step a quantum computer only allows 2-qubit gates between a limited set of qubit pairs, which is the coupling map or qubit connectivity graph. \tdok{Well, existing quantum processors do, but a general quantum computer has no such limitation. Here you are trying to say that you wish to run this on a real device, but this has never been said in the manuscript.}
\jk{Well the coupling map is not a limitation per se, if your coupling map is a clique, all 2-qubits combinations are possible, so no a general quantum computer do have a coupling map. And at the very beginning of the paper it is said that we do not design a algorithm for imaginary quantum computers but real ones.}

\begin{figure}[h]
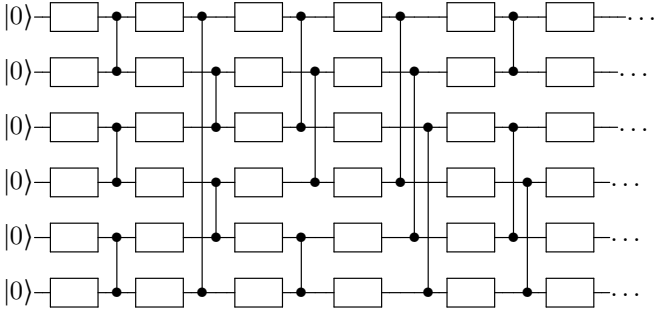

	\begin{center}
 \qcscaled{1}{
 	\ket0&&&\qw&\gate{{\color{MidnightBlue}\phantom{--}}}&\qw&\ctrl{1}&\qw&\gate{{\color{MidnightBlue}\phantom{--}}}&\qw&\ctrl{5}&\qw&\qw&\gate{{\color{MidnightBlue}\phantom{--}}}&\qw&\ctrl{2}&\qw&\qw&\gate{{\color{MidnightBlue}\phantom{--}}}&\qw&\ctrl{3}&\qw&\qw&\qw&\gate{{\color{MidnightBlue}\phantom{--}}}&\qw&\ctrl{1}&\qw&\qw&\gate{{\color{MidnightBlue}\phantom{--}}}&\qw&\qw&\qw&\qw&&\dots \\
	\ket0&&&\qw&\gate{{\color{MidnightBlue}\phantom{--}}}&\qw&\ctrl{0}&\qw&\gate{{\color{MidnightBlue}\phantom{--}}}&\qw&\qw&\ctrl{1}&\qw&\gate{{\color{MidnightBlue}\phantom{--}}}&\qw&\qw&\ctrl{2}&\qw&\gate{{\color{MidnightBlue}\phantom{--}}}&\qw&\qw&\ctrl{3}&\qw&\qw&\gate{{\color{MidnightBlue}\phantom{--}}}&\qw&\ctrl{0}&\qw&\qw&\gate{{\color{MidnightBlue}\phantom{--}}}&\qw&\qw&\qw&&\dots \\
	\ket0&&&\qw&\gate{{\color{MidnightBlue}\phantom{--}}}&\qw&\ctrl{1}&\qw&\gate{{\color{MidnightBlue}\phantom{--}}}&\qw&\qw&\ctrl{0}&\qw&\gate{{\color{MidnightBlue}\phantom{--}}}&\qw&\ctrl{0}&\qw&\qw&\gate{{\color{MidnightBlue}\phantom{--}}}&\qw&\qw&\qw&\ctrl{3}&\qw&\gate{{\color{MidnightBlue}\phantom{--}}}&\qw&\ctrl{2}&\qw&\qw&\gate{{\color{MidnightBlue}\phantom{--}}}&\qw&\qw&\qw&&\dots \\
	\ket0&&&\qw&\gate{{\color{MidnightBlue}\phantom{--}}}&\qw&\ctrl{0}&\qw&\gate{{\color{MidnightBlue}\phantom{--}}}&\qw&\qw&\ctrl{1}&\qw&\gate{{\color{MidnightBlue}\phantom{--}}}&\qw&\qw&\ctrl{0}&\qw&\gate{{\color{MidnightBlue}\phantom{--}}}&\qw&\ctrl{0}&\qw&\qw&\qw&\gate{{\color{MidnightBlue}\phantom{--}}}&\qw&\qw&\ctrl{2}&\qw&\gate{{\color{MidnightBlue}\phantom{--}}}&\qw&\qw&&\dots \\
	\ket0&&&\qw&\gate{{\color{MidnightBlue}\phantom{--}}}&\qw&\ctrl{1}&\qw&\gate{{\color{MidnightBlue}\phantom{--}}}&\qw&\qw&\ctrl{0}&\qw&\gate{{\color{MidnightBlue}\phantom{--}}}&\qw&\ctrl{1}&\qw&\qw&\gate{{\color{MidnightBlue}\phantom{--}}}&\qw&\qw&\ctrl{0}&\qw&\qw&\gate{{\color{MidnightBlue}\phantom{--}}}&\qw&\ctrl{0}&\qw&\qw&\gate{{\color{MidnightBlue}\phantom{--}}}&\qw&\qw&&\dots \\
	\ket0&&&\qw&\gate{{\color{MidnightBlue}\phantom{--}}}&\qw&\ctrl{0}&\qw&\gate{{\color{MidnightBlue}\phantom{--}}}&\qw&\ctrl{0}&\qw&\qw&\gate{{\color{MidnightBlue}\phantom{--}}}&\qw&\ctrl{0}&\qw&\qw&\gate{{\color{MidnightBlue}\phantom{--}}}&\qw&\qw&\qw&\ctrl{0}&\qw&\gate{{\color{MidnightBlue}\phantom{--}}}&\qw&\qw&\ctrl{0}&\qw&\gate{{\color{MidnightBlue}\phantom{--}}}&\qw&\qw&&\dots}\\\

			\end{center}
\caption{\label{fig:allcz} Alternating different entangling patterns on 6 fully connected qubits }
\end{figure}

Assuming that the qubit connectivity graph is connected, by choosing alternating entangling patterns (see Fig. \ref{fig:allcz}) it is possible in few steps to ensure that all qubits are entangled pairwise.
In designing our circuits, we will choose the succession of entangling patterns in such a way it minimizes the number of steps to reach a state where all qubits are entangled. Similar rules to design circuits are proposed in \cite{cerezo2020barren}.

Once the structure of the parametric circuit is fixed, the next decision we face is which parameters are going to be input parameters $\bm{\omega}$ (see section \ref{section:input}) and which are going to be model parameters $\bm{\theta}$.

Although the model parameters $\bm{\theta}$ have a similar role and are indistinguishable, each of the input parameters $\bm{\omega}$ corresponds to features that have a meaning, so their position in the circuit matters. This choice is today made experimentally on learning performance.

Also we implement the idea of \emph{data re-uploading}, suggested first by \cite{latorre2020} and used in \cite{lloyd2020qml}, which consists in inputting $\bm{\omega}$ more than once. Data re-uploading proves to be useful to add learning capacity without adding more qubits.

\newcommand{\ketone}{\ket{1}&&}
\newcommand{\ketplus}{\ket{+}&&}
\newcommand{\ketminus}{\ket{-}&&}
\newcommand{\qdots}{\dots&&}
\newcommand{\xygate}[1]{\measure{\!\scriptscriptstyle{#1}\!}}

\label{sec:experiments}
\section{Experimental protocol}

To assess the capabilities of our algorithm, we carry out a series of experiments on various classification problems:
the Iris flower dataset\cite{fisher1936iris},
the XOR problem with Gaussian noise,
the skin segmentation dataset\cite{dataSkin} and a four-classes synthetic dataset,
generated
using scikit-learn\cite{scikit-learn}.

In each case, we follow the same experimental procedure; first
we randomly split the dataset in a training and a testing subset, preserving the original ratio of each class.
Second, we use the training dataset to find the parameters~$\bm{\theta}^\star$ that minimize the loss function $\mathcal{L}_{\bm{\theta}}$.
Finally, we use the test set to evaluate the prediction capabilities on independent data of $\mathfrak{F}_{\bm{\theta}^\star}$ the trained model (see \cite[Ch. 7]{ESL}).

In the learning phase, we randomly initialize the parameters~$\bm{\theta}_0$ of the circuit and  optimize them iteratively as explained in section \ref{section:learning}. The loss function takes the full training data set as minibatch.

\newcommand{\bfgs}{{{BFGS}}}
\newcommand{\cobyla}{{{COBYLA}}}
We use the optimization package of SciPy \cite{2020SciPy}, with two different heuristics, namely the algorithms \bfgs\cite{byrd1995bfgs} and \cobyla\cite{cobyla1994}.

The \bfgs\ algorithm uses the gradient of the loss function $\mathcal{L}_{\bm{\theta}}$  and moves the parameters accordingly.
However, as $\mathcal{L}_{\bm{\theta}}$ depends on the random variable $\hat{P}_k$, it is random and its exact gradient is inaccessible. Nonetheless, with a high number of shots, the variance of $\hat{P}_k$ is low and we can resort to a finite differences estimate.

Moreover, for a small number of qubits, it is possible to simulate a quantum computer and get the theoretical $P_k$. We can then compute the \emph{exact} loss function using $P_k$ instead of its estimate.

On the other hand, as \cobyla\ algorithm is gradient-free, it can work with a low number of shots. Then, \cobyla\ is the preferred method using a real QPU  and \bfgs\ when using exact probabilities from a simulated QPU.

Once our quantum model is trained, we compare its results on the test set to those obtained using a classical model.
For this purpose, we consider the gradient boosting method XGBoost~\cite{xgboost} ---a state of the art model in many applications.

As a last note,  we fix the values introduced in section \ref{section:input}. In this paper, we  use an angular gap $\alpha\pi$ of $\frac{\pi}{10}$ and fix the quantile $q$ to 3, so we ignore at most $ \epsilon = 1\%$ of the points of every dataset\footnote{Iris: $\epsilon = 1\%$; Skin: $\epsilon = 0.8\%$; Artificial: $\epsilon = 0.5\%$.}.

\section{Iris dataset on QPU}\label{ssec:iris-exp}

\begin{figure}[ht]
	\centering
	\input{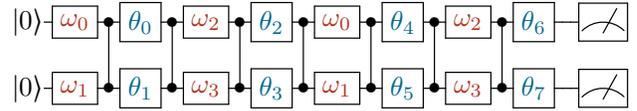}
	\vspace{1em}

	\caption{The quantum circuit for the classification of Iris dataset. Vector $\bm{\omega}$ encodes for (sepal length, sepal width, petal length, petal width). Note that the features are \emph{re-uploaded}.
    }
    \label{fig:iris-ibmq-circuit}
	\end{figure}

The Iris flower dataset\cite{fisher1936iris}, consists in 50 samples from each of three species of Iris (Iris setosa, Iris virginica and Iris versicolor), for a total of 150 data points, which we split in a train and test set of sizes 90 and 60 respectively.
The input vector of each sample consists of four features, the length and the width of  sepals and petals.

It is a well know test case for ternary classification, and, thanks to its relatively small sample size, we were able to run training and test entirely on an actual QPU, the {IBMq-valencia} 5-qubit quantum computer.

After trying several different circuits on simulated QPU, we choose to implement the four features in only two qubits.
Fig.~\ref{fig:iris-ibmq-circuit} shows our best performing circuit.
The three classes, setosa, virginica and versicolor, are read respectively in the bit strings $\mathsf{00}$, $\mathsf{01}$  and $\mathsf{10}$, where the leftmost bit is the measurement of the upmost qubit.

Since {IBMq-valencia} has five qubits, to halve the number of QPU calls, we parallelize computations by running two circuits at once on independent pairs of qubits, and we read the output accordingly.

At each learning iteration, to compute the loss function, we run the circuit for all 90 elements of the training set.
For the first 20 iterations the number of shots per run is 250, then we increase it to 500 and from iteration 50 we raise it again to 750 shots.

\begin{figure}[ht!]
	\centering
	\includegraphics[width=0.70\linewidth]{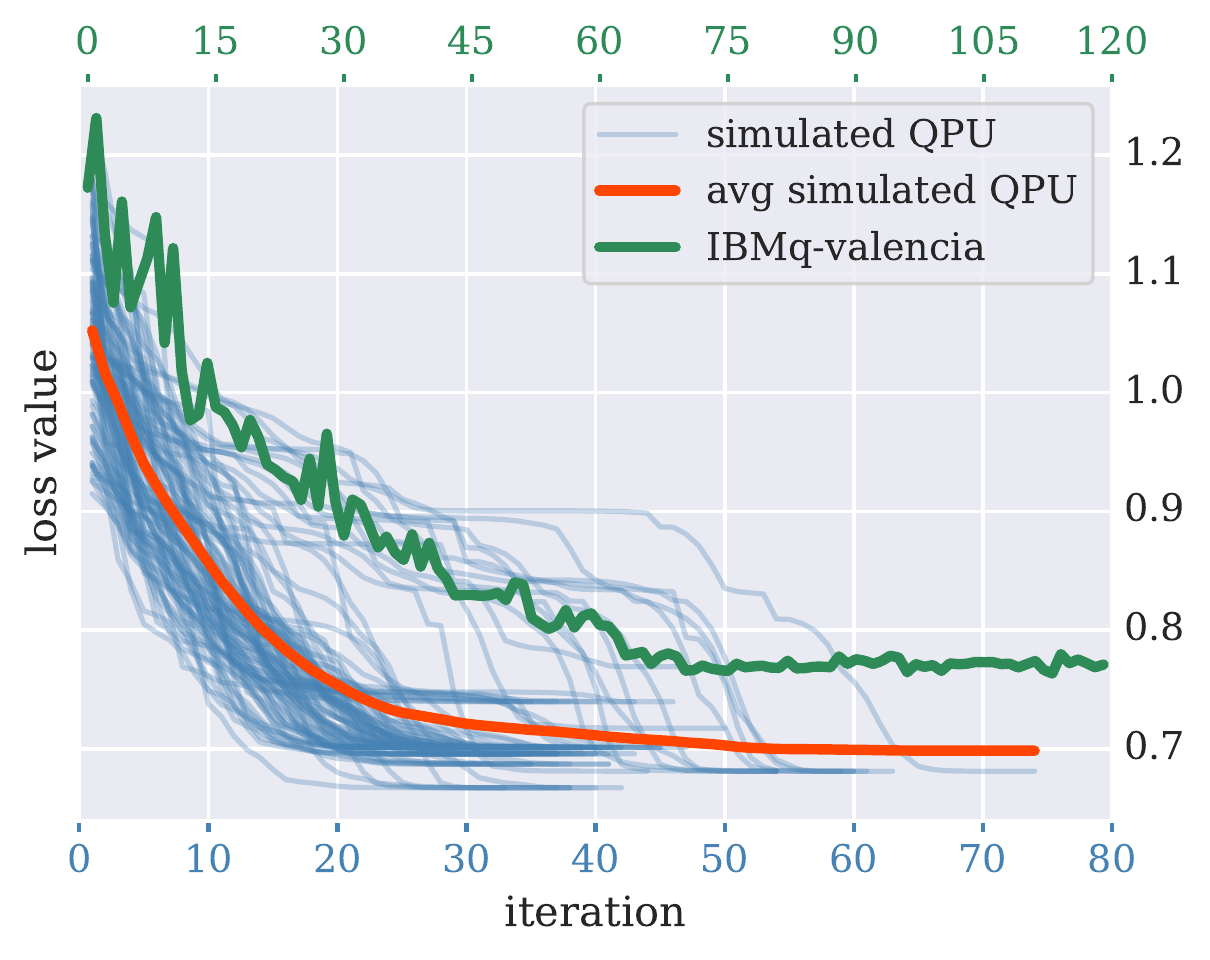}
	\caption{Evolution of the loss function value during training of the quantum Iris classifier. 
		The green line represents the training we did on {IBMq-valencia} with gradient-free optimizer \cobyla.
		Blue lines represent different trainings on simulated QPU using the exact loss function with gradient-based optimizer \bfgs, and red line shows their average.
		Note that the number of iterations for the {IBMq-valencia} training have their own scale (0-120), shown on top.
	}
	\label{fig:iris-loss-progress}
\end{figure}

We performed 120 optimization steps with \cobyla. Fig.~\ref{fig:iris-loss-progress} shows the evolution of the loss function, and we see that around iteration 80 it already stopped improving.
The experiment took 5400 dual circuit runs, for a total of $3.26$ millions shots, and spent approximately one hour and 20 minutes running on the QPU, as reported by IBM Quantum Experience platform.

We assess the resulting trained model over six different IBMq machines\footnote{essex, burlington, vigo, yorktown, london, ourense}, with three runs each and 300 shots per run.
Fig.~\ref{fig:iris-ibmq-tables} shows the confusion matrices and accuracies on the 60 points of the test set for both the quantum model and  the classical XGBoost model, trained with the same data.

\begin{figure}[h!]
	\centering
	\begin{minipage}{\columnwidth}
\begin{minipage}{.5\textwidth}
\centering \small \scshape
Quantum Model\\
{\footnotesize trained on IBMq-valencia}

\vspace*{.5em}
\begin{tabular}{l||r|r|r}
	& \bfseries 0 & \bfseries 1 & \bfseries 2 \\
	\hline \hline
	\bfseries 0 & $19.73$ & $0.27$ & $0.00$ \\
	\hline
	\bfseries 1 & $0.00$ & $19.93$ & $0.07$ \\
	\hline
	\bfseries 2 & $0.00$ & $2.40$ & $17.60$ \\
\end{tabular}

\vspace{.5em}

		Accuracy : $95.44 \pm 3.54 \%$

\end{minipage}
\begin{minipage}{.45\textwidth}
\centering \small \scshape
Classical Model\\
{\footnotesize ~}

\vspace*{.5em}
\begin{tabular}{l||r|r|r}
 & \bfseries 0 & \bfseries 1 & \bfseries 2 \\
\hline \hline
\bfseries 0 & 20 & 0 & 0 \\
\hline
\bfseries 1 & 0 & 19 & 1 \\
\hline
\bfseries 2 & 0 & 2 & 18 \\
\end{tabular}

\vspace{.5em}

Accuracy : $95.00\%$
\end{minipage}
\end{minipage}

	\caption{Compared test scores between our quantum model trained on IBMq \textit{valencia} and XGBoost a state-of-the-art classical model. We test the quantum model on six different IBMq machines with three runs each and the reported scores show the averages. In the confusion matrices, each entry $C_{i,j}$ is the number of observations actually in class~$i$ (row), but predicted to be in class~$j$~(column).
}
	\label{fig:iris-ibmq-tables}
\end{figure}

These results on quantum hardware are in line with the preliminary experiments with exact loss function on simulated QPU. We trained the quantum model a hundred times starting with different initial parameters and most of the time the training converges to close minimal values of the loss function (see Fig.~\ref{fig:iris-loss-progress}).
The learning process on IBMq-valencia however stagnated at higher values. Nonetheless,  the best model on simulated QPU, the one with the lowest training loss value, did not get a higher accuracy when tested on the six IBMq machines.

\section{Gaussian XOR on QPU}\label{ssec:xor-exp}

The XOR problem is a decision problem on binary input consisting in learning the exclusive-or function.
Given an input in $\{0,1\}^2$, the XOR function outputs 1 for $(0,1)$, $(1,0)$, and 0 for $(1,1)$, $(0,0)$.
Minsky and Papert \cite{minsky1969} showed that this problem is not linearly separable and cannot be learned by a perceptron\cite{perceptron1958}. 

\begin{wrapfigure}{r}{0.5\columnwidth}
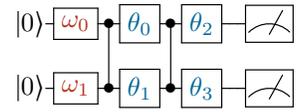

	\centering
		\qcscaled{1.}{\ketzero&\gate{{\color{BrickRed}\omega_{0}}}&\ctrl{0}\qwx[1]&\gate{{\color{MidnightBlue}\theta_{0}}}&\ctrl{0}\qwx[1]&\gate{{\color{MidnightBlue}\theta_{2}}}&\qw&\qw&\meter \\
            \ketzero&\gate{{\color{BrickRed}\omega_{1}}}&\ctrl{0}&\gate{{\color{MidnightBlue}\theta_{1}}}&\ctrl{0}&\gate{{\color{MidnightBlue}\theta_{3}}}&\qw&\qw&\meter}

	\caption{Circuit to solve the Gaussian XOR problem, vector $\bm{\omega}$ is the input.}
    \label{fig:xor-circuit}
\end{wrapfigure}

To make the problem more interesting and challenging, we extend the input space to continuous values and generate a dataset to train a model.
We take four symmetric points on the Cartesian axes, and assign label $0$ to the points on $x$-axis and label $1$ to those on $y$-axis
---this is an affine transformation of the original input points.
Then, we use these points as centers for four 2-dimensional Gaussian distributions, from each of which we sample 20 datapoints, assigning them the corresponding label.

\begin{figure}[h!]
		\centering
		\includegraphics[width=0.7\linewidth]{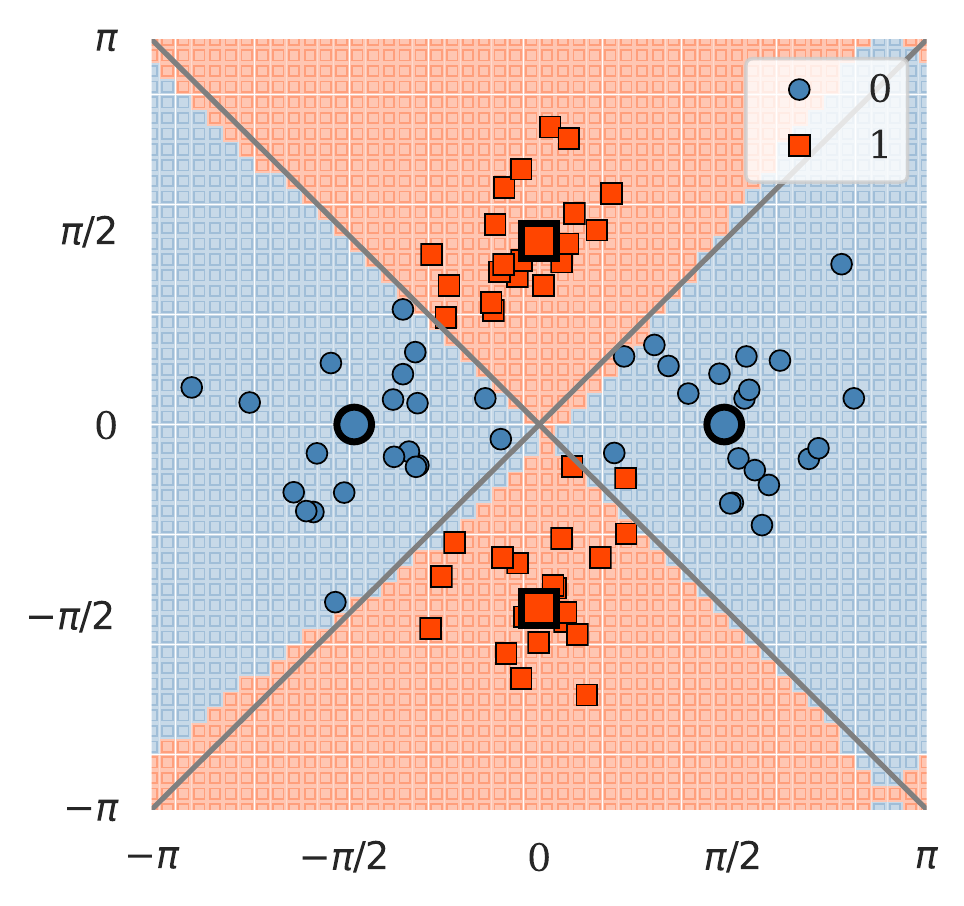}
	\caption{
		Scatter plot of the training dataset generated from the Gaussian XOR problem.
		The points with bold borders are the centers of the Gaussian distributions.
		The background shows, color coded, the predicted class w.r.t. $\bm{\omega}$, while the grey lines represent the Bayes-optimal decision boundaries. 
	}
	\label{fig:xor-pred}
\end{figure}

Figure~\ref{fig:xor-pred} shows the scatter plot of this generated dataset, along with the distribution centers.
We refer to this mixture of Gaussians as the \emph{Gaussian XOR} problem.

Our quantum classifier uses the two-qubit circuit with four parameters shown in Fig.~\ref{fig:xor-circuit}.
The datapoints coordinates are encoded in $\bm{\omega}$ without preprocessing.
We associate label $0$ with bit string $\mathsf{00}$ and label $1$ with bit string $\mathsf{10}$.

We train the model on {IBMq-valencia} with these 80 datapoints, proceeding as for the Iris dataset in section \ref{ssec:iris-exp}.
The optimization process converged after 69 steps, using \cobyla, for a total of 2760 dual circuit runs and $1.37$ million shots.

Since we know the generating distribution $\mathcal{D}$ of the dataset, we can compare our solution to the optimal-possible decision boundary.
This, is given by the  \emph{Bayes classifier} \cite[Sec. 2.4]{ESL}, which predicts the class directly from the posteriori distribution
%
$\hat{y} =\argmax_y P(y | \bm{\omega})$, where $P(y | \bm{\omega})$ is the conditional probability of label $y$ given observation  $\bm{\omega}$.


The generalization error of the Bayes classifier for this distribution is $3.33\%$, so the best possible accuracy is $96.67\%$.
Using one million newly generated datapoints we test our quantum model on a simulated QPU with 300 shots. The accuracy of this particular model ---trained with 80 points on actual hardware--- is $96.31\%$, not far from the theoretical maximum.

The decision boundaries of the Bayes classifier are the two perpendicular bisectors of the centers of the Gaussian distributions.
%
Figure~\ref{fig:xor-pred} shows the decision boundaries of our quantum model along with the Bayes-optimal ones.

\section{Skin segmentation dataset}\label{ssec:skin-exp}

The skin dataset consists in a large number of RGB-color values sampled from face images; each sample is labeled as skin or non-skin.
We use a random subset of size 1000, with equally represented classes, which we further separate in a test and train set containing 400 and 600 observations respectively.

\begin{wrapfigure}{r}{0.45\columnwidth}
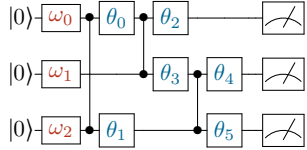

	\centering
		\qcscaled{0.87}{\ket0&&&\gate{{\color{BrickRed}\omega_{0}}}&\ctrl{0}\qwx[2]&\gate{{\color{MidnightBlue}\theta_{0}}}&\ctrl{0}\qwx[1]&\gate{{\color{MidnightBlue}\theta_{2}}}&\qw&\qw&\qw&\qw&\meter \\
			\ket0&&&\gate{{\color{BrickRed}\omega_{1}}}&\qw&\qw&\ctrl{0}&\gate{{\color{MidnightBlue}\theta_{3}}}&\ctrl{0}\qwx[1]&\gate{{\color{MidnightBlue}\theta_{4}}}&\qw&\qw&\meter \\
			\ket0&&&\gate{{\color{BrickRed}\omega_{2}}}&\ctrl{0}&\gate{{\color{MidnightBlue}\theta_{1}}}&\qw&\qw&\ctrl{0}&\gate{{\color{MidnightBlue}\theta_{5}}}&\qw&\qw&\meter}
	\caption{\label{fig:skin-circuit}The circuit for skin segmentation quantum model. Vector $\bm{\omega}$ encodes for (B,G,R).}
\end{wrapfigure}

\begin{figure}[h!]
	\centering
	\begin{minipage}{.75\columnwidth}
		\begin{minipage}{0.49\columnwidth}
			\includegraphics[width=1\columnwidth]{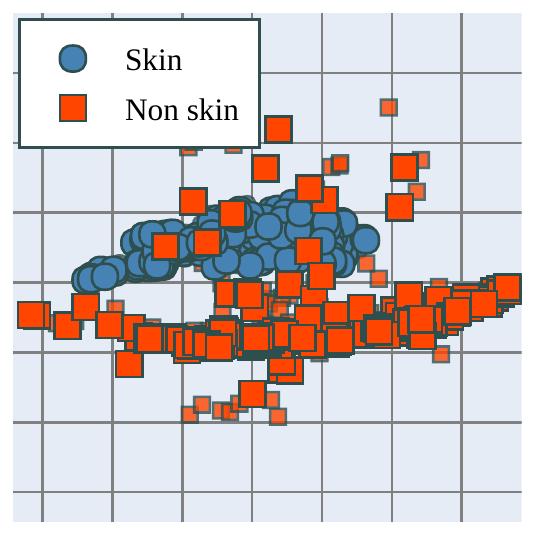}
		\end{minipage}
		\begin{minipage}{0.49\columnwidth}
			\includegraphics[width=1\columnwidth]{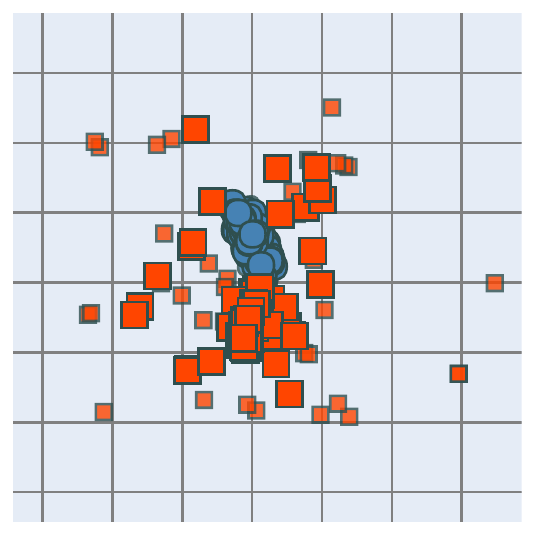}
		\end{minipage}
	\end{minipage}
	\caption{\label{fig:skin-pca}Skin segmentation dataset seen from two orthogonal angles.}
\end{figure}

The dataset is not linearly separable (see Fig. \ref{fig:skin-pca}) and we want to know if our algorithm could do the binary classification with a minimalist circuit.
After several attempts we manage to get good results with the 3-qubit circuit in Fig. \ref{fig:skin-circuit}, with no data re-uploading and only 3 entanglements and 6 parameters.
We associate the classes with the permutation-invariant bit strings $\mathsf{000}$ and $\mathsf{111}$.
We train and test the model with the simulated QPU using exact probabilities. The test set consists of 400 samples, with 200 examples per class. Our simple quantum model on 3-qubit scored a 94\% accuracy, 
close to the 98\% of the classical XGBoost model.

\section{Synthetic dataset with 4 classes}\label{ssec:synthetic}

In previous sections, we treated two real-life datasets, which are well known examples of binary and ternary classification.
Nonetheless, our algorithm can, in principle, discriminate as many classes as  bit strings the circuit can output. So a 2-qubit circuit should be able to solve a quaternary classification problem.

Then, we generate a 4-class bidimensional synthetic dataset of 5000 samples using
the \texttt{make\_classification} function from
 scikit-learn~\cite{scikit-learn}.
We  randomly split the dataset in train and test sets representing  respectively $60\%$ and $40\%$ of the samples, preserving each class ratio.

\begin{figure}[h!]
    \centering
    \input{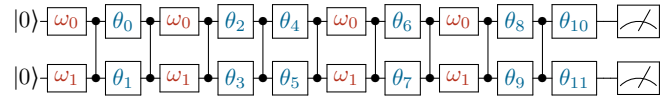}

    \caption{The quantum circuit for quaternary classification of the two-dimensional synthetic dataset.
    }
    \label{fig:artificial-circuit}
\end{figure}

Out of many 2-qubit circuits, we choose a circuit with twelve parameters and four data uploadings as defined in Fig.~\ref{fig:artificial-circuit}.
We assign each of the possible bit strings --$\mathsf{00}$, $\mathsf{01}$, $\mathsf{10}$, $\mathsf{11}$-- to a class.

We train and test this quantum model using exact probabilities on simulated QPU.
The average accuracy over a hundred experiments with different initial parameters is $85\pm 2.8\%$, 
close to the $87.98\pm 0\%$ of the classical XGBoost model.

\begin{figure}[t]
	\centering
	\includegraphics[width=.75\linewidth]{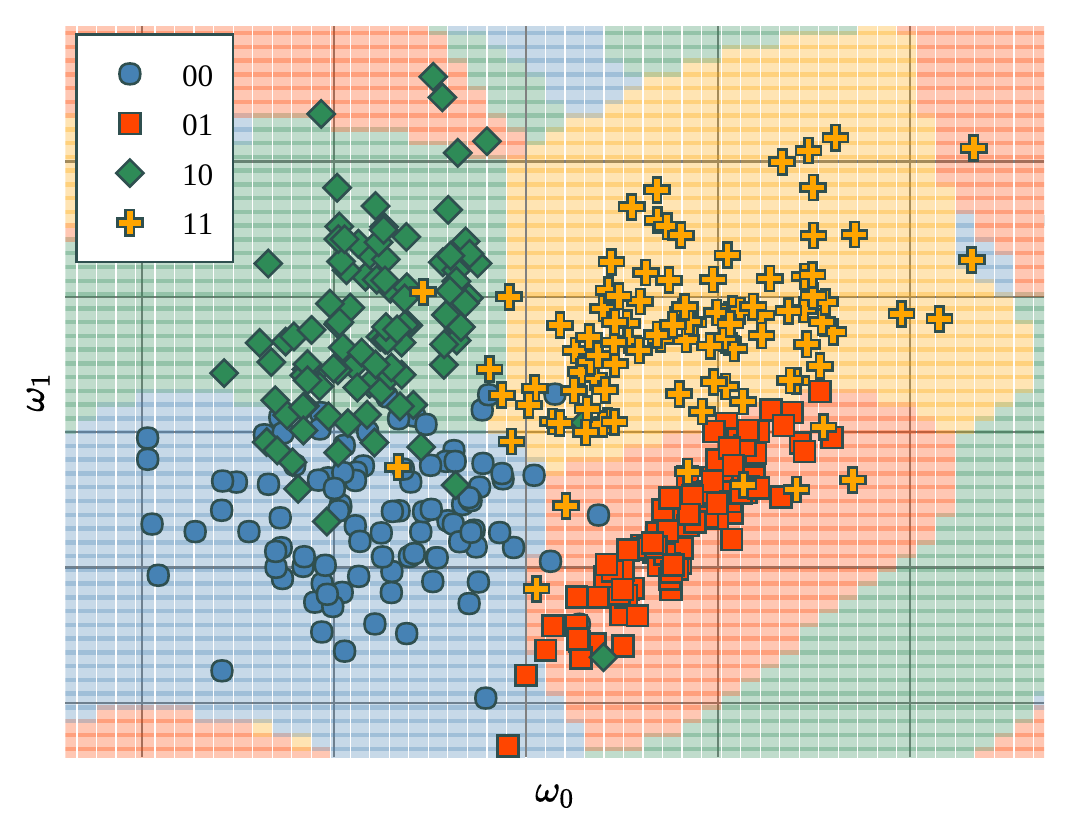}

	\caption{\label{fig:quaternary-decision} Scatter plot of the synthetic dataset. The background shows, color coded, the predicted class w.r.t. $\omega_0$ and $\omega_1$.}
\end{figure}

Fig.~\ref{fig:quaternary-decision} shows the prediction for the whole feature space of one the trained models.
We note non trivial decision boundaries.

	\section{Future work and conclusion}
We obtain satisfying quantum models for the challenging Iris flower dataset and the other polyadic problems, with good test scores compared with a classical model.

Furthermore, we were able to train the model for the Iris dataset and the Gaussian XOR problem on actual hardware. This was possible because of the low number of shots needed and the noise resilience of the model. Moreover, even though different machines have different noise patterns, the model trained on one machine performed well on all others (see Fig.~\ref{fig:iris-ibmq-tables}).

It is important to remind that the optimization heuristics used in this work have severe limitations. The loss function is quantum computed and thus random; hence, we can only approximate its gradient to some extent, at the expense of more shots. Usual gradient based methods as \bfgs\ perform poorly with this sort of randomness. In fact, we use them successfully on simulations, but cannot do the same on actual hardware. For QPUs we resort to the gradient-free \cobyla. However, it is well known\cite{powell2007view} that \cobyla's performance drastically degrades with the dimension of the parameter space.

To improve the algorithm, we will need to explore optimization techniques adapted to quantum loss functions. One salient idea is to use circuit modifications to approximate the gradient, as proposed by \cite{crooks2019gradients}\cite{schuld2018gradients}.

Another avenue worth exploring is to work in a different parameter space. Here, we optimize directly the angles $\bm{\theta}$ of the parametric quantum circuit; instead, we could train the model using an intermediate parameter vector  $\bm{\rho}$ mapped to~$\bm{\theta}$. For instance, a non-linear mapping would reshape the optimization landscape. Also, this would allow constraints on~$\bm{\theta}$, in particular parameter sharing, as in convolutional neural networks\cite{goodfellow2016deeplearning}.

However, when addressing higher dimensional data and increasing the number of qubits, the most challenging issue becomes circuit design. We cannot rely anymore on sheer guess-and-check and need to understand better why some circuits outperform others.

As the computational capability of quantum hardware keeps increasing, we expect the empirical study of more complex quantum machine learning models to give new insights and lead to increasingly better algorithms.


\appendices
	\section{Translating to hardware specific native gates}
\label{native}

Quantum hardware manufacturers implement different sets of native gates to specify the circuits to run on their machines. In section \ref{section:optimizing} we define our own set of gates to specify circuits: the 0-pulse $Z$-rotation and the 1-pulse $s_x$ one-qubit gates and the controlled-$Z$ two-qubit gate.

The one-qubit gates are native in IBM\cite{qiskit}, Rigetti\cite{pyquil2016rigetti}, Honeywell\cite{honeywell2020} and Google\cite{cirq} quantum architectures and implemented with the same number of qubits.

The controlled-$Z$ is native for Rigetti's QPU and for most of Google's ---with the notable exception of the Sycamore QPU. The tunable two-qubit gate\cite{supremacy} of the Sycamore processor has not been studied in this paper.

Honeywell uses the two-qubit $Z\!z$ gate and $\cz_j^i$ translates in  $Z^i_{-\pi/2}Z^j_{-\pi/2}{Z\!z}_j^i$ which preserves the number of pulses.

IBM uses the controlled-not as two-qubit gate, noted $C^i_j$.  Controlled-$Z$ can be translated in $H_jC^i_jH_j$, where $H$ is the Hadamard gate. With the following translations
\begin{eqnarray*}
H s_x Z_\phi s_x H &=& s_x Z_{\pi-\phi} s_x \\
H s_x Z_\phi s_x &=& Z_\pi s_x Z_{\phi-\frac{\pi}{2}} s_x \\
s_x Z_\phi s_x H &=& s_x Z_{\phi-\frac{\pi}{2}} s_x Z_\pi
\end{eqnarray*}
the Hadamard gates disappear, thus preserving the number of pulses.

%
%
%

\bibliographystyle{IEEEtran}
\bibliography{biblio.bib}

\end{document}